\documentclass{elsart}

\usepackage{graphicx}

\usepackage{amssymb}

\begin{document}

\begin{frontmatter}



\title{The Spin-Boson model with a structured environment:  A
comparison of approaches}


\author[LMU]{F.K.\ Wilhelm\corauthref{cor1}}
\ead{wilhelm@theorie.physik.uni-muenchen.de}
\corauth[cor1]{Corresponding author} \author[LMU]{S.\ Kleff}
\author[LMU]{J.\ von Delft} \address[LMU]{Sektion Physik and CeNS,
Ludwig-Maximilians-Universit\"at,  Theresienstr. 37, D-80333
M\"unchen, Germany}

\begin{abstract}
In the spin boson model, the properties of the oscillator bath  are
fully  characterized by the spectral density of oscillators
$J(\omega)$. We study the case when this function is of Breit-Wigner
shape and has a sharp peak at a frequency $\Omega$ with width
$\Gamma\ll\Omega$. We use  a number of approaches such as the
weak-coupling Bloch-Redfield equation, the non-interacting blip
approximation (NIBA)  and the flow-equation renormalization scheme. We
show, that if $\Omega$ is  much larger than the qubit energy scales,
the dynamics corresponds to an Ohmic spin boson model with a strongly
reduced tunnel splitting. We also show that the direction of the
scaling of the tunnel splitting  changes sign when the bare splitting
crosses $\Omega$. 
We find good
agreement between our analytical approximations and numerical
results. We illuminate how and why different approaches to the model
account for these features and discuss the interpretation of this
model in the context of an application to quantum computation and
read-out.
\end{abstract}

\end{frontmatter}

\section{Introduction}
\label{}

The subject of open-system quantum mechanics and the physics of the
boundary between classical and quantum physics has been of strong
interest since the early days of quantum theory. A paradigmatic
standard model for the study of open quantum systems is the spin boson
model\cite{Leggett,Weiss}: A single two-state system coupled to a bath
of harmonic oscillators described by the Hamiltonian
\begin{equation}
H=\frac12(\epsilon\hat{\sigma}_z+\Delta\hat{\sigma}_z)+\frac12
\hat{\sigma}_z\sum_i c_i
x_i+\frac12\sum_i\left(\frac{p_i^2}{m}+m_i\omega_i^2x_i^2\right)+H_0.
\label{eq:hamiltonian}
\end{equation}
Here, $H_0$ is a constant counter-term. The energy eigenvalues of the
two state system alone are $\pm E/2$ with
$E=\sqrt{\epsilon^2+\Delta^2}$.  The oscillator bath can model
arbitrary {\em Gaussian} noise sources. It is fully characterized by a
spectral function $J(\omega)$ which depends on the distribution of
frequencies and couplings
\begin{equation}
J(\omega)=\sum_i \frac{c_i^2}{2m_i\omega_i^2}\delta(\omega-\omega_i).
\end{equation}
For a given physical system, e.g.\ a superconducting quantum bit
coupled to a noisy electronic circuit, $J(\omega)$ can be obtained by
analyzing either the effective friction \cite{ASSP} or noise
\cite{EPJB} originating from the environment. It is useful to also
introduce the semiclassical noise power
$S(\omega)=J(\omega)\coth(\omega/2T)$.

Next to its long tradition in chemical physics, the physics of open
quantum systems and in particular the spin boson model has gained
recent practical  importance in the field of quantum
computation\cite{Chuang}. There,  one is interested in obtaining long
phase coherence times for the actual  computation and long relaxation
times for the readout. In a number of  quantum computation
realizations, the researcher has the option to taylor or engineer at
least part of the properties of the quantum system and the
dissipative environment under study \cite{Engineering}, e.g.\ in the case of
superconducting qubits coupled to their control and readout
electronics\cite{EPJB}.  In particular, environements with nontrivial
internal dynamics, e.g.\ with resonances, can  be realized and appear
to be attractive \cite{EPJB,Neumann,Martinis}. Much is known about the
physics of the spin boson model whose spectral density  is a power law
with an exponential cutoff \cite{Leggett,Weiss}. Such spectral
densities only contain the cutoff as an energy scale, which is
typically assumed to be very high, leading to scale-free results.

Much less is known about structured environments.  We are interested
in a generic realization of this physics described by a spectral
density containing a Breit-Wigner resonance
\begin{equation}
J(\omega)=\alpha\omega
\frac{\Omega^4}{(\omega^2-\Omega^2)^2+4\omega^2\Gamma^2}
\label{eq:jomega}
\end{equation}
in the underdamped case $\Gamma\ll\Omega$. In that case, we can expand
\begin{equation}
J(\omega)=\frac{\alpha\Omega^3}{8i\Gamma}\sum_{\sigma,\sigma^\prime=\pm1}
\frac{\sigma\sigma^\prime}{\omega-i\sigma^\prime\tilde{\Omega}-i\sigma\Gamma}\quad\quad
\tilde{\Omega}=\Omega-2\Gamma^2/\Omega.
\label{eq:residue_repr}
\end{equation}
Moreover, we will be able to profit from the analytical continuation
 of results with  a Drude spectral density, noting that eq.\
 \ref{eq:jomega} can be written as
\begin{equation}
J(\omega)=\frac{\alpha\Omega^3}{4i\Gamma}\sum_{\sigma=\pm1}
\frac{\sigma\omega}{\omega^2-(\sigma\tilde{\Omega}+i\Gamma)^2}.
\end{equation}
Note, that the shift of the resonance position from  $\Omega$ towards
$\tilde{\Omega}$ can be neglected in the underdamped case, except
close to the resonance. Hence, we will henceforth only emphasize this
shift in those cases, when it  actually affects the results.

This type of spectral density is generically obtained by coupling the
spin to a harmonic oscillator with eigenfrequency $\Omega$ which in
turn  is damped with a linear friction coeffcient $\Gamma/2$. This
friction is modeled quantum-mechanically by a bath of harmonic
oscillators.  Using a normal mode transformation, one can show that
this is equivalent to our spin boson Hamiltonian with a structured
bath \cite{Garg}. More details are given in  section
\ref{sec:discussion}. This model is realized in various physical systems
such as chemical reactions involving biomolecules \cite{Garg}, atoms
in cavities \cite{CavityQED} or superconducting qubits coupled to
resonators \cite{EPJB,Martinis,Zorin,Blais,Marquardt,Plastina}.  It corresponds to the
nonlinear dimer model of polaron physics \cite{Kenkre}. The case of no
dissipation with restriction to a rotating wave approximation is known
in quantum optics as the Jaynes-Cummings model\cite{QuantenOptik}.
Our notation corresponds to the one adopted in Ref.\ \cite{Neumann}
and is slightly different to the one of Ref.\ \cite{KleffKehrein}.

The spin-boson model cannot be solved exactly and  has been studied by
a number of approaches. Some of them are largely numerical such as
quantum Monte Carlo \cite{Egger}, real-time renormalization group
\cite{Schoeller}, quasiadiabatic path integrals \cite{Thorwart} flow
equation renormalization \cite{Kehrein} and numerical renormalization
group \cite{Costi}, others are mainly  analytical such as the
noninteracting blip approximation (NIBA), a systematic weak damping
approximation or exact Born approximation \cite{Paladino,Loss} or
Bloch-Redfield \cite{Redfield,Agyres,Hartmann}.  A spectral density of
the type eq.\ \ref{eq:jomega} poses a challenge to most of these
approaches, since the spectral density in units of the frequency,
$J(\omega)/\omega$, can be either very small (off-resonance) or large
(on-resonance). In order to explore the phyiscs of this model and to
obtain useful analytical  information, these approximation schemes
have to be applied within their range of validity and compared to
numerical methods which are essentially nonperturbative in
$J(\omega)/\omega$. Alternatively, one can 
treat the coupled TSS and oscillator system as multilevel quantum system 
and only the friction to the oscillator as a bath \cite{ThorwartHere}. 

The plan of this paper is to analyze this model using the
weak-coupling Bloch-Redfield theory and the nonperturbative NIBA and
to compare the results to a full numerical study obtained in the flow equation
scheme. We will  very briefly introduce these methods and compare the
dynamics of the reduced density matrix [ characterized through the
expecation value $s_z(t)=\left\langle\hat{\sigma}_z\right\rangle (t)$
with, for definiteness,  localized initial condition $s_z(0)=1$],
effective reduced Hamiltonians, dephasing and coherence
rates. Interpreteations of the results in terms of a superconducting
quantum bit coupled to a resonant measurement circuit will be given.

\section{Bloch-Redfield}

The Bloch-Redfield-theory has originally been developped in the
context of nuclear magnetic resonance\cite{Redfield}. It offers a
systematic way to obtain a generalized master equation within the weak
coupling Born approximation between system and bath with
$J(\omega)/\omega$ as small parameter. It contains a subtle Markov
approximation such that the resulting master equation
is local in time; however, the main bath correlations relevant within
the Born approximation are kept and they do lead to time-dependent
rates for a driven system \cite{Hartmann,Marlies}. Bloch-Redfield has been shown to be numerically
equivalent to a full non-Markovian path integral technique for a rather generic
choice of parameters \cite{Hartmann}. Nevertheless, recent calculations 
at $T=0$ seem to
indicate \cite{Loss}  that there may under certain circumstances 
be additional terms in
the Born approximation, that are neglected in the Bloch-Redfield approach.

The natural starting point for the Bloch Redfield theory in the
undriven case are the energy eigenstates of the spin-part of the
Hamiltonian  \ref{eq:hamiltonian}. In that ``energy basis'',  the
Bloch-Redfield equation can be  written as (see e.g.\ 
ref.\ \cite{Weiss})
\begin{equation}
\dot{\rho}_nm=-i\omega_{nm}\rho_{nm}+\sum_{kl}R_{nmkl}\rho_{kl}
\end{equation}
where all indices take the values $+$ and $-$ corresponding to the
ground and excited state and $\omega_{nm}=(E_n-E_m)/\hbar$. 
The Redfield tensor has the form
\begin{equation}
R_{nmkl}=\delta_{lm}\sum_r \Gamma_{nrrk}^{(+)}+\delta_{nk}
\sum\Gamma_{lrrm}^{(-)}-\Gamma_{lmnk}^{(+)}-\Gamma_{lmnk}^{(-)}
\label{eq:redfield_tensor}
\end{equation}
where we have introduced
\begin{equation}
\Gamma_{lmnk}^{(+)}=(\sigma_z)_{lm}(\sigma_z)_{nk}\Gamma(\omega_{nk})
\quad{\rm and}\quad
\Gamma_{lmnk}^{(-)}=(\sigma_z)_{lm}(\sigma_z)_{nk}\Gamma^\ast(-\omega_{lm})
\label{eq:golden_rates}
\end{equation}
where $(\sigma_z)_{nk}$ are matrix elements of $\sigma_z$ in the
energy basis, the $^\ast$ denotes complex conjugation.  The basic
building block of the rates in the Redfield tensor is the rate
$\Gamma$ which can be written as
\begin{equation}
\Gamma(\delta\omega)=\frac{1}{2\pi\hbar}\int_0^\infty dt
e^{-i\delta\omega t}e^{-0^+t}\int_0^\infty d\omega  J(\omega)\left[\cos\omega
t\coth(\omega/2T)-i\sin\omega t\right].
\label{eq:br_basicrate}
\end{equation}
The
resulting dynamics displays exponential decay and reads 
\begin{equation}
s_z(t)=\frac{\epsilon^2}{E_{\rm eff}^2}\left(e^{-\Gamma_r
t}+\tanh\left(\frac{E_{\rm eff}}{2T}\right)(1-e^{-\Gamma_r t})\right)
+\frac{\Delta_{\rm eff}^2}{E^2_{\rm eff}}\cos(E_{\rm eff}
t)e^{-\Gamma_\phi t}.
\label{eq:br_dynamics}
\end{equation}
The quantities $\Delta_{\rm eff}$ and 
$E_{\rm eff}=\sqrt{\epsilon^2+\Delta_{\rm eff}^2}$ can be associated with 
a renormalized Hamiltonian
\begin{equation}
H_{\rm
eff}=\frac12\left(\epsilon\hat{\sigma}_z+\Delta_{\rm eff}\hat{\sigma}_z\right).
\label{Heff}
\end{equation}
The details of the shift of the tunnel splitting 
$\Delta\mapsto\Delta_{\rm
eff}$ will be
discussed below.

The $\Gamma_r$ term in eq.\ \ref{eq:br_dynamics} describes incoherent
energy relaxation. It leads the system into thermal
occupation of the renormalized Hamiltonian described below. The
relaxation rate can be deduced from the Bloch-Redfield rates eqs.\
\ref{eq:br_basicrate},  \ref{eq:redfield_tensor} and
\ref{eq:golden_rates}
\begin{eqnarray}
\Gamma_r&=&R_{----}+R_{++++}=(\sigma_z)_{-+}(\sigma_z)_{+-}
\left(\Gamma(E)+\Gamma(-E)+c.c.\right)\nonumber\\
&=&\frac{\Delta^2}{2E^2}S(E).
\label{eq:br_relax}
\end{eqnarray}
This result is easily understood in terms of the Born-approximation:
In order to relax, the system has to exchange the energy corresponding
to the energy splitting $E$ 
with the
environment at once, using a single photon.

The last term in eq.\ \ref{eq:br_dynamics} describes
quantum coherent oscillations analogous to Larmor precession of a spin
\cite{Slichter}.  These are the hallmark of (macroscopic) quantum
coherence in the spin boson system. Their decay rate can hence be
identified with the dephasing rate and can, using  eqs.\
\ref{eq:br_basicrate},  \ref{eq:redfield_tensor} and
\ref{eq:golden_rates}, be written as
\begin{eqnarray}
\Gamma_\phi &=&-{\rm Re} \Gamma_{-+-+}={\rm
Re}\left[2(\sigma_z)_{--}(\sigma_z)_{++}
\Gamma(0)+(\sigma_z)_{-+}(\sigma_z)_{+-}(\Gamma(E)+\Gamma^\ast(-E))\right]
\nonumber\\ &=&\frac{\Gamma_r}{2}+\alpha T
\frac{\epsilon^2}{\Delta^2}.
\label{eq:br_dephasrate}
\end{eqnarray}
Note, that on very general grounds \cite{Slichter} we have
$2\Gamma_\phi\ge\Gamma_r$. The extra factor of $1/2$ originates from the
fact that there are in principle {\em two} dephasing channels corresponding
to clockwise and counterclockwise Larmor precession. We are following
here the standard NMR-motivated notation \cite{Slichter}; one could
equivalently define $2\Gamma_\phi$ as the true physical dephasing
rate. The first term in eq.\ \ref{eq:br_dephasrate} is proportional
to the relaxation rate eq.\ \ref{eq:br_relax}, which reflects that a
relaxation process certainly also randomizes the phase information. 
The additional term involves $S(0)$, which in our case is $\propto T$.
This contribution originates from 
``flipless'' dephasing processes which randomize the phases
while keeping energy constant, i.e.\ transitions from a state into
itself.

The form of both rates eqs.\ \ref{eq:br_dephasrate} and \ref{eq:br_relax} 
resembles the case of unstructured enviroments \cite{Paladino},
even though the spectral density eq.\ \ref{eq:residue_repr}  has
singularities close to the real axis. The high relaxation rate at 
$E\simeq\Omega$ corresponds to
resonant interaction between the qubit and the central environmental
oscillator. When interpreting this result, one has to 
be aware, that the Born
approximation involved is only valid for $\Gamma_{r,\phi}\ll E$,
which, bounding $J(\omega)\le J(\Omega)$, means 
$\alpha \Omega^2<\Gamma^2$. This
a very rigorous constraint in the underdamped case, $\Gamma\ll\Omega$, which
 we are considering. Also
physically, we do not expect this result to be consistent up to strong
couplings, because the relatively weakly damped big oscillator is a
highly coherent quantum system which mostly {\em reversibly} exchanges
energy with the spin. However, since the Golden-Rule approximation  in
Bloch-Redfield only takes the long time limit, this reversible
exchange cannot be seen in the Bloch-Redfield result.  
This can be understood from
the  order of limits prescribed by Bloch-Redfield and shown in the
Appendix: The imaginary part of the energy is {\em first} sent to
zero. Non-Markovian approximation schemes \cite{Paladino,Loss} would
at least  take a self-consistent  value and thus shift the $S(E)$ in
eq.\ \ref{eq:br_relax} into ${\rm Re} S(E+i\Gamma_r)$.  Such shifts
can be important in particular if $E\simeq\Omega$, when both predicted
rates are very high.
\begin{figure}[tb]
\includegraphics[width=0.99\columnwidth]{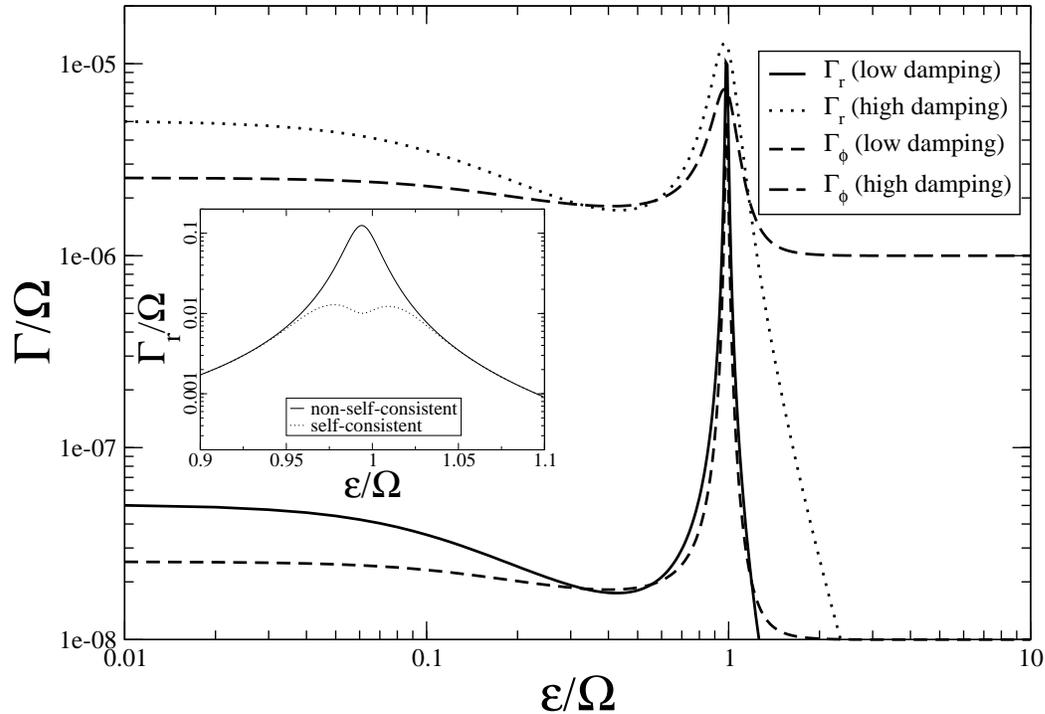}
\caption{Relaxation and dephasing rates as a function of energy bias
predicted from Bloch-Redfield theory. Parameters are
$\Delta=0.1\Omega$, $T=0.01\Omega$.  For the low-damping plots we have
chosen $\alpha=10^{-6}$, $\Gamma=10^{-2}\Omega$, for the high-damping
plots we have chosen $\alpha=10^{-4}$, $\Gamma=10^{-4}\Omega$. The
inset compares the self-consistent and non-self-consistent relaxation
rate around the resonance for $\Delta=0.1\Omega$, $T=0$,
$\alpha=10^{-2}$ and $\Gamma=10^{-2}\Omega$.
\label{fig:blochrates}}
\end{figure}
Results are summarized in figure \ref{fig:blochrates}. We clearly see
the peaked behavior at resonance and notice that the
influence of the self-consistent solution is rather small 
even at rather extreme
parameters. On the other hand, the self-consistent solution predicts lower
rates as compaed to the non-self-consistent one, similar to the predictions
of flow-equation studies \cite{KleffKehrein}.

As mentioned above, the environment not only causes dephasing
and relaxation, it also renormalizes the tunnel splitting
$\Delta$ (and with it the transition frequency), by dressing the
two-state system with  environmental degrees of freedom. This is
similar to the physics of the Lamb shift or the Franck Condon effect
and leads, in the nonperturbative regime, to the 
dissipative quantum 
phase transition \cite{Leggett,Chakravarty,Schmid}. In our
case,  the transition frequency is renormalized according to
$E\rightarrow E-{\rm Im} R_{+-+-}$. If we look at the imaginary part
of the generic rate, eq.\ \ref{eq:br_imagrate},
$\Gamma^\prime(E)=\frac{1}{4\pi\hbar}\int d\omega\;  J(\omega){\mathcal
P}\frac{1}{\omega^2-E^2}\left[\coth(\beta \omega/2)E-\omega\right]$
we observe a weight
function ${\mathcal P} (\omega^2-E^2)^{-1}$  which changes sign at
$\omega\simeq E$. Thus we can expect an upward renormalization
 of $E$ if most of
the spectral weight of $J(\omega)$ is above $E$ (corresponding  to
$E<\Omega$) whereas $E$ scales downward in the opposite case. Physically,
this corresponds to level repulsion between the spin and the oscillators
in the environment. The result also
is
consistent with usual second order perturbation theory for the
energies. The sign changes happens at $E\simeq\Omega$, the point where 
most of the spectral weight is concentrated, thus we expect a rather
sharp structure of the splitting $E_{\rm eff}(\Omega)$. 
Note, that this sign change is not predicted for the usual spin-boson 
in the scaling limit, which can be 
studied
by the well-known adiabatic
renormalization approach\cite{Leggett,Chakravarty}. In that case,
$E_{\rm eff}$ is always reduced. This is consistent
with our findings, because in the scaling limit, the vast majority of the
environmental oscillators have high frequency, much above the qubit splitting.

\begin{figure}[tb]
\includegraphics[width=0.99\columnwidth]{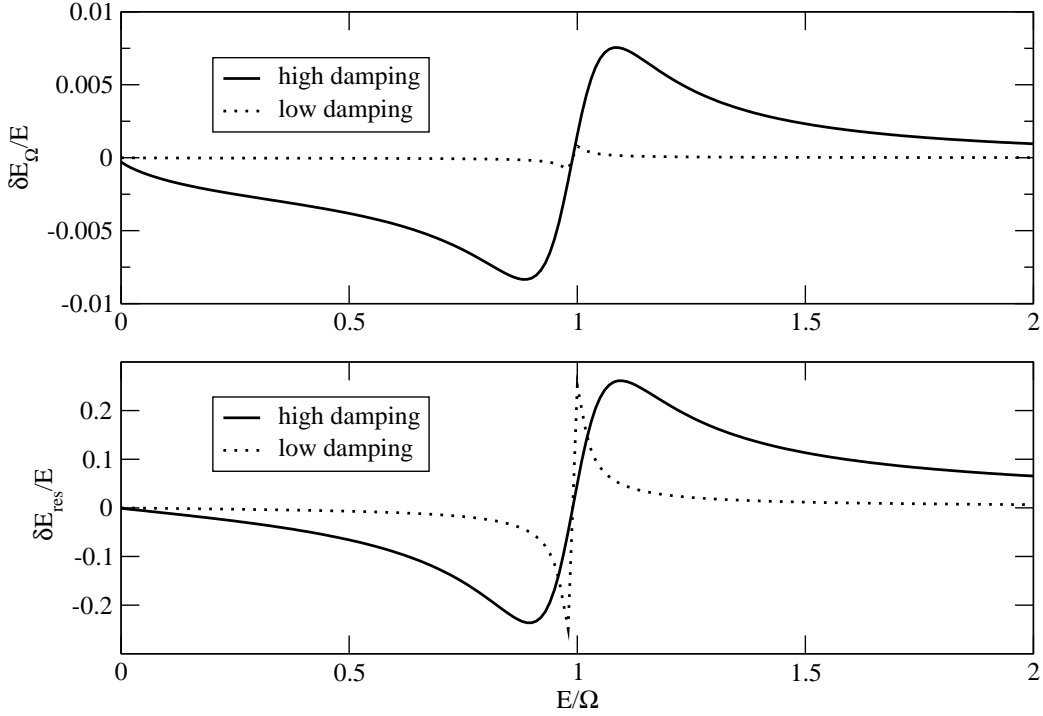}
\caption{Renormalization of the energy splitting at $T=0$ taken  at
the degeneracy point $\epsilon=0$, defined as {\em positive} if the
splitting is decreased. Upper panel: Ohmic-like logarithmic
contribution from eq.\ \ref{eq:br_renormal_ohm}; lower panel: Contribution of the environmental resonance from eq.\ \ref{eq:br_renormal_peak}
(for discussion see text). Low damping: $\Gamma=0.01\Omega$,
$\alpha=10^{-4}$, high damping: $\Gamma=0.1\Omega$, $\alpha=10^{-2}$.
 \label{fig:br_renormal}}
\end{figure}

From the structure of the dephasing rate eq.\ \ref{eq:br_dephasrate}
we can conclude that the last term in eq.\ \ref{eq:br_imgamma}, which
is even in energy, drops from the final result.  Moreover, the
remaining contribution to eq.\ \ref{eq:br_imgamma} vanishes as
$E\rightarrow 0$. If we finally go to low temperatures, we can replace
$p$ in eq.\ \ref{eq:br_imgamma} by an appropriate logarithm and find
for the shift of the transition frequency
\begin{equation}
\delta E=\frac{\Delta^2}{E^2}\frac{\alpha}{2\pi}\frac{i\Omega^2
E}{\Gamma} \sum_\sigma
\frac{\sigma}{E^2-(\sigma\tilde{\Omega}^2+i\Gamma)^2}\log\left(\frac{\Gamma-i\sigma\tilde{\Omega}}{iE}\right).
\label{eq:br_renormal_full}
\end{equation}
In the underdamped limit we are working in, we can  approximate the
logarithm as $\log|\Omega/E|-i\sigma\pi/2$ and split the result as
$E_{\rm eff}=E+\delta E$,  $\delta E=\delta E_\Omega+\delta E_{\rm
res}$.  It contains a logarithmic contribution which resembles the
scaling in the Ohmic case (with cutoff frequency $\Omega$),
\begin{equation}
\delta E_\Omega=\frac{2}{\pi}\frac{\Delta^2}{E^2}J(E)\log
\left|\frac{E}{\tilde{\Omega}}\right|.
\label{eq:br_renormal_ohm}
\end{equation}
This contribution changes sign from an upward shift at $\Omega<E$ to a
downward shift at $\Omega>E$ as is expected from the general
arguments above. The logarithmic divergence at
low $E$ can be indicated as a precursor of a dissipative phase
transition. The other contribution takes into account the enormous
spectral weight of the resonance,
\begin{equation}
\delta E_{\rm
res}=\frac{\Delta^2}{E^2}J(E)\frac{E^2-\tilde{\Omega}^2}{\Omega\Gamma}.
\label{eq:br_renormal_peak}
\end{equation}
This contribution is of the order $\alpha/\Gamma$. It will be shown below, 
that terms of this kind persist even in the absence of damping of the 
external oscillator. It, too, 
undergoes the expected sign change. It is linear at low $E$ and
hence does not contribute to a dissipative phase transition. It instead
represents a substantial but finite renormalization. This is due to 
the fact, that for a dissipative phase transition, the environment has to
get entangled with the spin down to arbitrarily small frequencies.  
These results are
summarized in  figure \ref{fig:br_renormal}.
As expected, we find in fig.\ \ref{fig:br_renormal}, that the 
energy shift has a sharp structure around the sign change at $E\simeq\Omega$. 
At this point, the spin becomes strongly
entangled with the central oscillator, hence the concept of ``qubit 
energy splitting'' is of limited applicability. 
This observation is consistent with the usual dressed atom approach
of cavity quantum electrodynamics.

\section{NIBA}

So far, we have restricted ourselves to the Born approximation, i.e.\
to the lowest order in $J(\omega)/\omega$ and have otherwise kept the
system general. We now turn to the noninteracting blip approximation
(NIBA), which is nonperturbative in that parameter.  It can be derived
from evaluating the influence functional in a path-integral approach
by assuming that the off-diagonal excursions (``blips'')  of the
density matrix contributing to the path of the two state system are
uncorrelated \cite{Leggett,Weiss}.  It is thus justified when 
$E\ll\Omega$, because then the bath is oscillating rapidly on the scale
of the two-state system and the time-integrated bath correlation function
quickly averages out, leading to weak damping on longer time scales. 
Alternatively, the NIBA can be obtained by
analyzing a the polaron-transfomed version of the spin-boson
Hamiltonian.

NIBA is known to work well under these conditions 
at the degeneracy point $\epsilon=0$.  At
$\epsilon\not=0$, the situation is more subtle. 
At 
$\epsilon\gg\Delta$ \cite{Leggett,Weiss} the true dynamics is dominated
by incoherent relaxation, which is again accurately predicted.
This application of 
NIBA is closely related to the so-called $P(E)$ theory of Coulomb blockade
\cite{Weiss,IngoldNazarov,Gerd,WZS}.

In this approach, the dynamics turns out to be governed by
the Laplace transformed exponentiated correlation function
\begin{equation}
P(\lambda)=\frac{\Delta^2}{2\pi} \int_0^{\infty} e^{-\lambda
t}e^{K(t)}dt
\label{eq:niba_plambda}
\end{equation}
where $K(t)$ is the twice integrated bath correlation function from eq.\
\ref{eq:br_basicrate}
\begin{equation}
K(t)=\frac{1}{2\pi}\int_0^\infty dt \frac{J(\omega)}{\omega^2}\left((\cos\omega t-1)
{\rm coth}(\omega/2T) +i\sin\omega t\right).
\label{eq:niba_kt}
\end{equation}
At the degeneracy point, the dynamics of the system in Laplace space
is readily found from
\begin{equation}
s_z(\lambda)=\int_0^\infty e^{-\lambda t}s_z(t)
=\frac{1}{\lambda+{\rm Re} S(\lambda)}
\label{eq:niba_dynamics}
\end{equation}
where $S(\lambda)=(P(\lambda)+P^\ast(\lambda^\ast))/2$.  Far from the
degeneracy point, we find incoherent relaxation
\begin{equation}
s_z(t)=e^{-\Gamma_r
t}\left[1-\tanh\left(\frac{\epsilon}{2T}\right)\right]
+\tanh\left(\frac{\epsilon}{2T}\right)\quad\quad \Gamma_r=2{\rm Re}
P(i\epsilon+0)
\label{eq:niba_relax}.
\end{equation}

\begin{figure}[tb]
\quad\includegraphics[width=0.80\columnwidth]{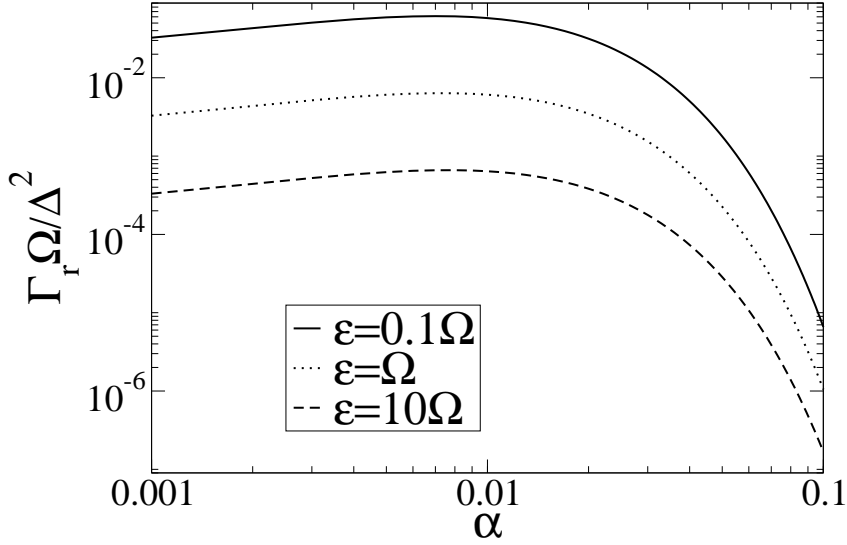}
\caption{Relaxation rates calculated from the NIBA 
eq.\ \ref{eq:niba_relax_long}, in the long time
approximation using $\Gamma=0.01\Omega$. Note, that for small
$\alpha$ values the rate first grows with growing $\alpha$,  until the
localization due to the resonance takes over and relaxation rates drop
dramatically.
 \label{fig:niba_relaxation}}
\end{figure}

\begin{figure}[tb]
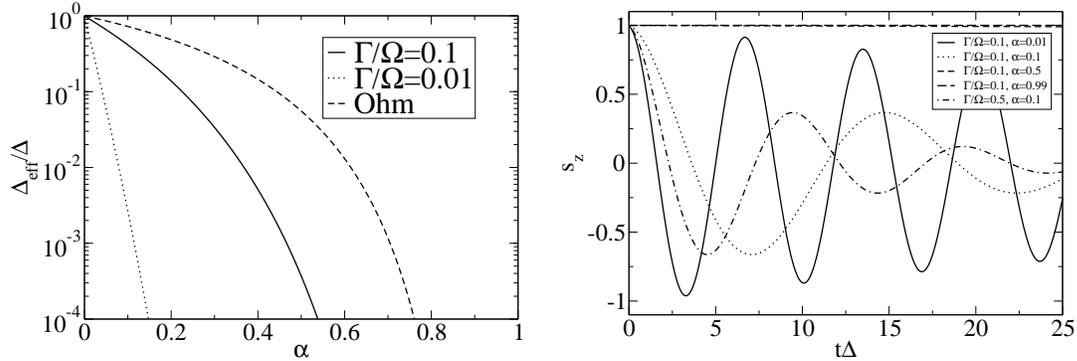

\includegraphics[width=0.49\columnwidth]{niba_deff2.eps}
\quad\includegraphics[width=0.49\columnwidth]{niba_dyn_long2.eps}
\caption{Dynamical properties within the long time approximation of
  the NIBA at $\epsilon=0$.  
Left panel: Effective tunnel matrix element eq.\ \ref{eq:niba_slowdown} for
  $\Omega/\Delta=10$ for different values of the environment linewidth
  as a function of $\alpha$.  Lower $\Gamma$ means more spectral
  weight and stronger downscaling of $\Delta_{\rm eff}$. Right panel:
  Expectation value of $\sigma_z$, eq.\ \ref{eq:mittag} for different values of the damping
  parameters and $\Delta/\Omega=0.1$.
 \label{fig:niba_dynamic_long}}
\end{figure}

At $T=0$, we can use eq.\ \ref{eq:residue_repr} to 
evaluate $K(t)$ in closed  form
\begin{eqnarray}
K(t)&=&\frac{\alpha\Omega}{8\pi\Gamma i}\sum_{\sigma\sigma^\prime}
\frac{\sigma^\prime\Omega-i\Gamma\sigma}{i\sigma^\prime\Gamma+\sigma\Omega}
\left[e^{(i\sigma\Omega-\sigma^\prime\Gamma)t}{\rm
Ei}(-(i\sigma\Omega-\sigma^\prime\Gamma)t)\right.\nonumber\\
&&\left.-\gamma-\log(-i(\sigma\Omega-\sigma^\prime \Gamma)t) \right]
\label{eq:niba_ktT0}
\end{eqnarray}
This is too compliated to allow a direct calculation of
 $P(\lambda)$ from
eq.\ \ref{eq:niba_plambda}. 
At low energies, $E\ll\Omega$
 we can concentrate on the long time limit of  eq.\
\ref{eq:niba_ktT0} and find, keeping only lowest order terms in
$\Gamma/\Omega$
\begin{equation}
K_{\rm long}(t)=-\frac{\alpha\Omega}{\Gamma}
-2\alpha\left[\log|\Omega t|+\gamma+i\pi/2\right]
\end{equation}
where $\gamma$ is the Euler-Mascheroni-constant. 
This is a combination of a constant term of the order
$\alpha/\Gamma$ and a logarithmic term which resembles the findings in the 
Ohmic case \cite{Leggett,Weiss}. This is similar to
what we observed in our Bloch-Redfield result for the scaling in 
eqs.\ \ref{eq:br_renormal_ohm} and \ref{eq:br_renormal_peak}.  From
here, we find $P(\lambda)$ being
\begin{equation}
P(\lambda)=e^{-\alpha\Omega/\Gamma}e^{-2\gamma\alpha}e^{-i\alpha\pi}
\frac{\Delta^2}{\lambda}\Gamma(1-2\alpha)\left(\frac{\lambda}{\Omega}\right)^{2\alpha}.
\label{eq:niba_plong}
\end{equation}
Off the degeneracy point, we can directly evaluate the relaxation rate
from eq.\ \ref{eq:niba_relax} which reads
\begin{equation}
\Gamma_r=\frac{\Delta^2}{|\epsilon|}e^{-\frac{\alpha\Omega}{\Gamma}}\frac{e^{-2\gamma\alpha}}{\Gamma(2\alpha)}\left(\frac{|\epsilon|}{\Omega}\right)^{2\alpha}
\label{eq:niba_relax_long}.
\end{equation}
This rate resembles to the Ohmic case
\cite{IngoldNazarov,Weiss} but is reduced by an extra exponential
prefactor $\exp(-\alpha\Omega/\Gamma)$, which again represents the
contribution of the resonance and can be very small. Thus, we find te
important result that by
desigining small $\alpha$ but appreciable $\alpha\Omega/\Gamma$, the
incoherent relaxation rate of the spin can be reduced to extremely small
values. A physical interpretation of this finding will be given
later on. The predictions of equation \ref{eq:niba_relax_long} are
shown in fig.\ \ref{fig:niba_relaxation}.
At the degeneracy point, at $\epsilon=0$, we find the Laplace
transform of $s_z$ using eqs.\ \ref{eq:niba_dynamics} and
\ref{eq:niba_plong}. In analogy to the Ohmic case \cite{Weiss,Leggett}
we obtain for the backtransform that
\begin{equation}
s_z(t)=E_{2-2\alpha}\left(-(\Delta_{\rm eff} t)^{2-2\alpha}\right)
\label{eq:mittag}
\end{equation} where $E$ is
the  Mittag-Leffler function \cite{Weiss,Leggett,Erdelyi} and
\begin{equation}
\Delta_{\rm
eff}=\Delta\left(\frac{\Delta}{\Omega}\right)^{\alpha/(1-\alpha)}
\left(e^{-\alpha\Omega/\Gamma}e^{-2\gamma\alpha}\cos\pi\alpha\Gamma(1-2\alpha)\right)^{1/(2-2\alpha)}
\label{eq:niba_slowdown}
\end{equation}
is the renormalized tunnel splitting. Note that this is only valid at
$\Delta\ll\Omega$ because we have taken the long time limit for
$K(t)$. Consequently, it always predicts a downward renormalization. 
As in the Ohmic case, the dynamics show a crossover from decaying
oscillations at low $\alpha$ to  incoherent decay at $\alpha\ge
1/2$ at $\epsilon=0$. The renormalized tunneling frequency 
$\Delta_{\rm eff}$ shows a
combination of the usual Ohmic scaling behavior governed  by $\alpha$,
including a dissipative phase transition at $\alpha=1$, plus a very
effective downscaling of $e^{-\alpha\Omega\pi/8\Gamma(1-\alpha)}$
governed by $\alpha/\Gamma$ only, which also occurs for an undamped 
resonance and
is not present for the Ohmic case. This again captures the
contribution of the resonance and  reflects the behavior we have
observed in eqs.\ \ref{eq:br_renormal_peak}  and fig.\
\ref{fig:br_renormal} to lowest order in $\alpha/\Gamma$.  The
dynamics is illustrated in figure \ref{fig:niba_dynamic_long}. We can 
observe, that the time evolution of the spin is almost brought to 
a standstill, in the sense of absence of {\em both} oscillations and
decay, already at modest coupling constants.

\section{Comparison to flow equation results}

\begin{figure}[tb]
\includegraphics[width=0.99\columnwidth]{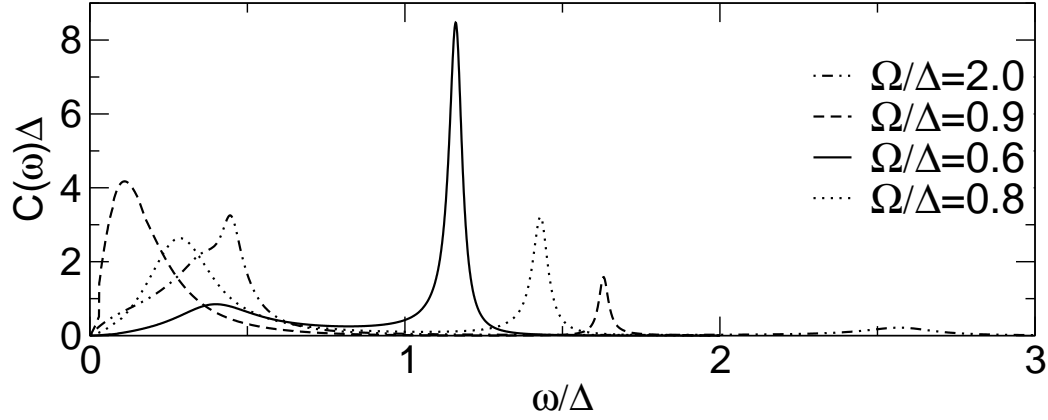}
\caption{Correlation function as evaluated from the flow-equation
method  using $\Omega/\Gamma=0.06$ and $\alpha=0.15$.
 \label{fig:flow_corr}}
\end{figure}

\begin{figure}[tb]
\includegraphics[width=0.99\columnwidth]{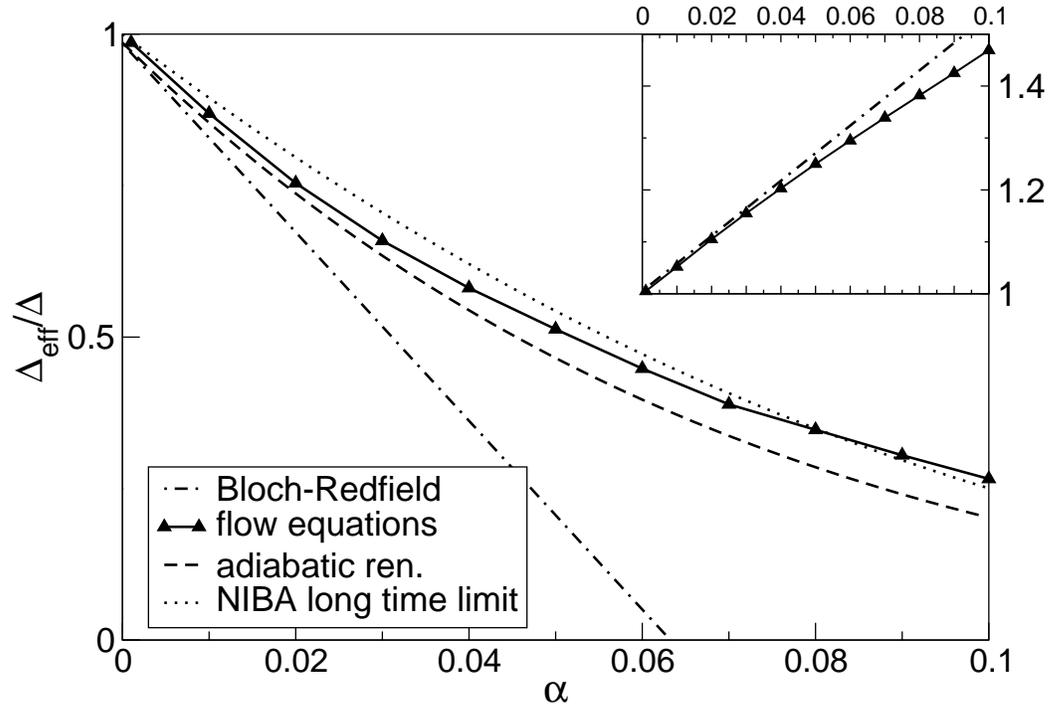}
\caption{Rescaling of the tunneling matrix element using different
  methods.  Main plot: $\Gamma/\Omega=0.02\pi $, $\Delta/\Omega=0.1$;
inset: $\Gamma/\Omega=0.06\pi $, $\Delta/\Omega=1.1$.
 \label{fig:adiabatic}}
\end{figure}

So far, we have studied our system using traditional methods for
open quantum systems. In order to complement this work, we want  to
compare the above results with previous work \cite{KleffKehrein},
in which the same setup was studied with the flow-equation 
renormalization method
\cite{Kehrein}, which originates from strongly correlated
electron systems and very well suited for treating problems with
several different energy scales. We will restrict ourselves to $\epsilon=0$. 
This method
typically can be used to calculate spin-spin correlation functions
in equilibrium such as $C(t)=\langle\sigma_z(t)\sigma_z(0)\rangle_{\rm eq}$. A typical
example is shown in fig.\ \ref{fig:flow_corr}.  The Fourier-transformed 
 correlation
function $C(\omega)$ is peaked at several frequencies.  
The resonance around
$\Delta_{\rm eff}$ corresponds to coherent oscillations, its width can
be identified with
the dephasing rate. There can also be a resonance around $\Omega$ 
corresponding to oscillations of the oscillator
leaving a trace on the qubit, but it hardly carries spectral weight.
We have numerically solved the flow equations for small and moderate
coupling strengths. More complete results are published elsewhere
\cite{KleffKehrein}.  We see, that at $\Delta\ll\Omega$, $\Delta_{\rm
eff}$ is rescaled downwards similar to the NIBA, but 
 with quantitative differences
The Bloch-Redfield result produces the correct slope at small $\alpha$, 
see fig.\ \ref{fig:adiabatic}. Around
$\Delta=\Omega$, the rescaling changes sign. Remarkably,
Bloch-Redfield also predicts the slope above the 
sign change with good accuracy, see inset of fig.\ \ref{fig:adiabatic}, although
this set of date is taken very close to resonance. 
Please note, that  in the inset
fig.\ \ref{fig:adiabatic} it is important
to keep
$\tilde{\Omega}$
in eq.\ \ref{eq:br_renormal_full}.


\section{Relation to quantum measurement and 
entanglement\label{sec:discussion}}

As already mentioned in the introduction, a straightforward way to
implement this model with the spectral density eq.\ \ref{eq:jomega} is to
couple  the TSS to a single harmonic oscillator with resonance
frequency $\Omega$, which is in turn damped by additional
oscillators. This model has the  Hamiltonian
\begin{eqnarray}
\hat{H}&=&\frac{\epsilon}{2}\hat{\sigma}_z+\frac{\Delta}{2}\hat{\sigma}_x
+\frac{\hat{P}^2}{2M}
+\frac{M}{2}\Omega^2(\hat{X}-q\hat{\sigma}_z)^2\nonumber\\
&&+\sum_i\left(\frac{\hat{p}_i^2}{2m_i}
+\frac{m_i}{2}\omega_i^2(\hat{x}_i-(\tilde{c}_i/m_i\omega_i^2)\hat{X})^2\right).\label{eq:althamiltonian}
\end{eqnarray}
The oscillator bath is characterized through an ohmic spectral density
$ \tilde{J}(\omega)=\sum \frac{\pi
\tilde{c}_i^2}{2m_i\omega_i}\delta(\omega-\omega_i) =M\Gamma\omega, $
where, $\Gamma/2$ is the friction coefficient of the damped big
oscillator.  It was shown in \cite{Garg}, using a normal-mode
decomposition, that this system is equivalent to the spin-boson
Hamiltonian eq.\ \ref{eq:hamiltonian} with spectral density eq.\
\ref{eq:jomega}, where $\alpha=2Mq^2\Gamma/\hbar$.

\begin{figure}
\includegraphics[width=0.99\columnwidth]{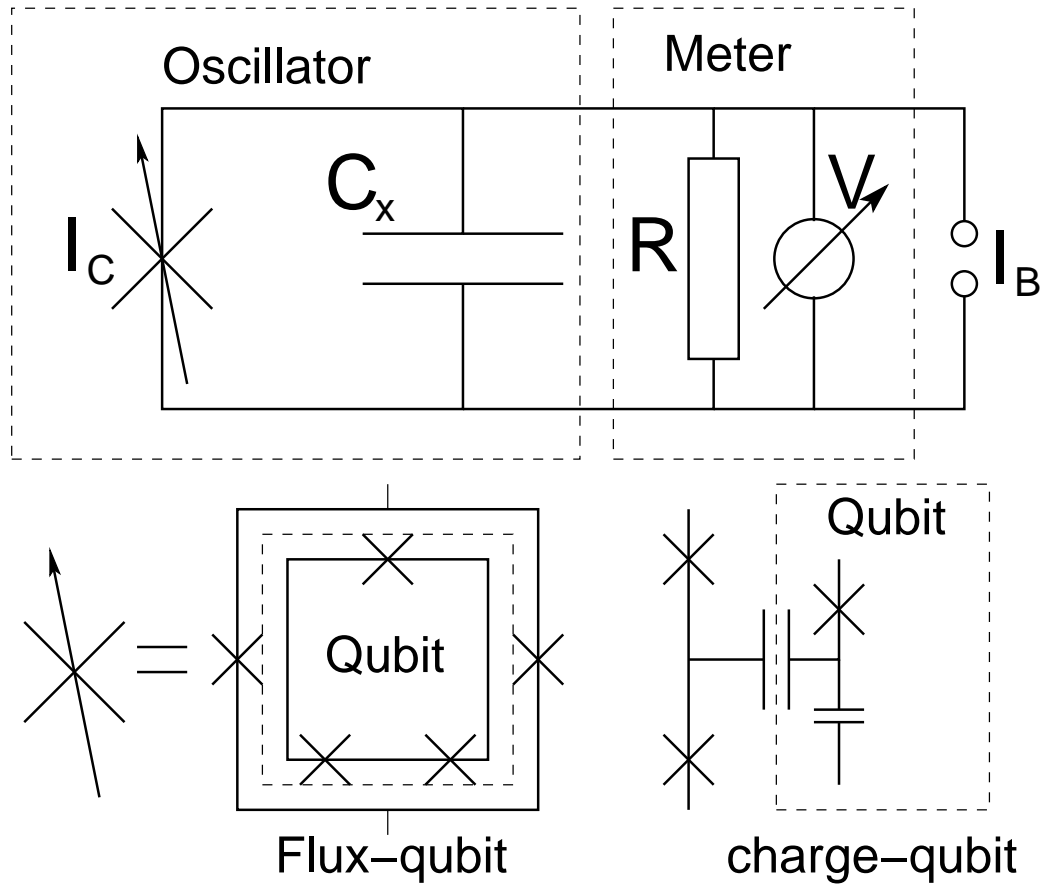}
\caption{Underdamped read-out devices for superconducting flux (left)
and charge (right) quantum bits \label{fig:circuit} involving a
tunable Josephson junction representing a SQUID or a superconducting
SET.}
\end{figure}

\begin{figure}
\includegraphics[width=0.99\columnwidth]{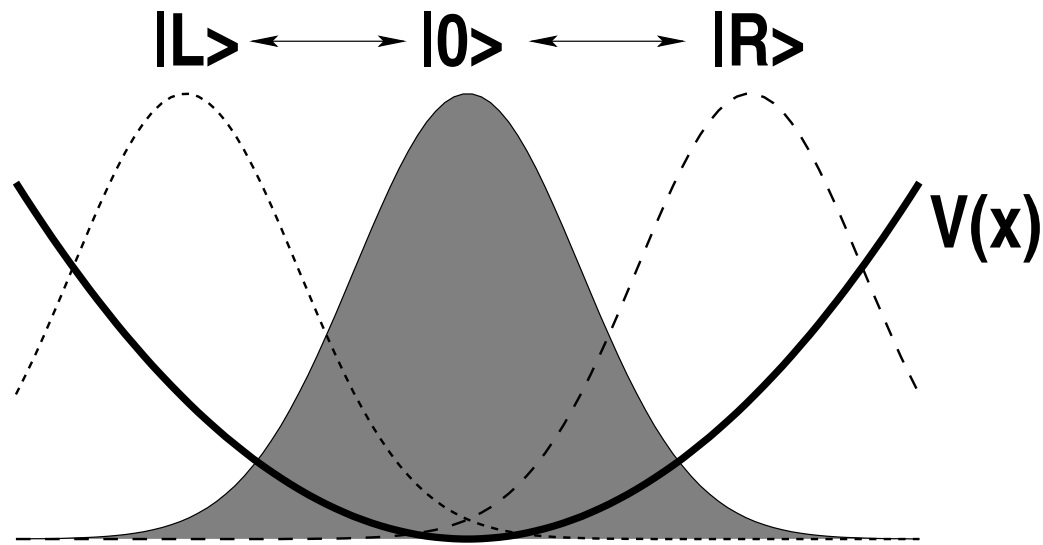}
\caption{Visualization of the ground state $|0\rangle$ and the
coherent pointer-states $|L\rangle$ and $|R\rangle$ of the oscillator
\label{pointers} in the potential $V(x)$}
\end{figure}

There are a number of realizations of such models. We would like to
concentrate on a realization in superconducting quantum circuits: A
flux quantum bit coupled to the plasma resonance of a DC-SQUID. This
setup has been thoroughly analyzed in Refs.\ \cite{ASSP,EPJB}. It has
been shown that the spectral density of the flux noise indeed leads
to eq.\ \ref{eq:jomega} and how the circuit parameters relate to the
parameters of that function. Moreover, it has been shown that the
coupling parameter $q$ actually can be tuned by the bias current
through the SQUID. A  representative circuit is shown in
fig. \ref{fig:circuit}. It is also shown there and discussed in Ref.\
\cite{Neumann}, that a similar though less favorable realization can
be found for charge quantum bits. We are mentioning this model,
because it describes a detector of a quantum variable. Thus, we are
going to interpret the findings of this paper in terms of quantum
measurement theory. Other applications of resonantors coupled to
superconducting qubits have been discussed in \cite{Blais,Marquardt,Plastina}
As a key result, we have found above within Bloch-Redfield as well as
within NIBA, that at $\epsilon,\Delta\ll\Omega$, the system dynamics
can be interpreted as an Ohmic spin boson model with a strongly
downscaled tunneling matrix  element. This can be understood in terms
of the following model, which was introduced and discussed already in
Ref.\ \cite{Neumann}. We start from the undamped case, $\Gamma=0$.
the low-energy Hilbert space is spanned by $|\pm\rangle_{\rm
eff}=|\pm\rangle|L/R\rangle$ where $|\pm\rangle$ are the basis states
of the qubit,  $\sigma_z|\pm\rangle=\pm|\pm\rangle$ and $|L/R\rangle$
are coherent states of the harmonic oscillator centered around $X=\pm
q$, see Fig.\ \ref{pointers}.  So in a general low-energy state
$|\psi\rangle=a|+\rangle_{\rm eff}+b|-\rangle_{\rm eff}$,
$|a|^2+|b|^2=1$,  qubit and oscillator are entangled.  In this
low-energy basis, the Hamiltonian acquires the form of the
renonrmalized spin part of the spin boson Hamiltonian
eq. (\ref{Heff}), with $\Delta_{\rm eff}=\Delta\langle
L|R\rangle=\Delta e^{-\eta}$, where $\eta=M\Omega q^2/\hbar$.  This
coincides with the result of eq.\ \ref{eq:niba_slowdown} in the limit
of $\alpha\rightarrow0$ but $\alpha/\Gamma={\rm const.}$  Under an
appropriate choice of parameters, we can achieve $\eta>1$ and
$\Delta_{\rm eff}\ll\Delta$. Following the notion of  Ref.\
\cite{Cirac}, the degree of entanglement is equal to
$1-e^{-2\eta}=1-|\Delta_{\rm eff}/\Delta|^2$, i.e.\ we can interpret
strong separation of the preferred states of the external oscillator
and strong renormalization,i.e.\ $\Delta_{\rm eff}/\Delta\ll1$ with strong entanglement.
In terms of quantum measurement theory, the oscillator states are
pointers onto the qubit states \cite{Zurek}. Chosing $\eta\gg1$ corresponds
to the condition of almost  {\em orthogonal} pointer states in the
environment, which has been identified as the condition for an ideal
detector-dominated von-Neumann-measurement\cite{Zurek,Peres}. Such a
measurement corresponds to  the standard textbook quantum measurement:
The preferred states into which superpositions are decohered are
assumed to be Eigenstates of the measured observable regardless of the Hamiltonian
of the qubit. In our case, eq.\ \ref{eq:althamiltonian} describes
coupling of the pointer degree of freedom to $\hat{\sigma}_z$ and
hence measurement thereof. Rescaling $\Delta_{\rm eff}$ asymptotically
to zero means bringing the target states of the decoherence
arbitrarily close to Eigenstates of $\hat{\sigma}_z$, thus
realizing the aforementioned textbook assumption.

As it stands, the qubit just gets entangled with the pointers, but
they are not read out. This can be done by coupling to the dissipative
environment.  As shown above, its influence corresponds to that of an
Ohmic environment of strength $\alpha$. Taking $\alpha\ll1$, this
leads to dephasing and relaxation rates analogous to the Bloch
Redfield results eqs.\ \ref{eq:br_relax} and \ref{eq:br_dephasrate}
\begin{equation}
\Gamma_{\rm r}=\pi\alpha\frac{\Delta_{\rm eff}^2}{E_{\rm eff}}{\rm
coth} \left(\frac{E_{\rm eff}}{2T}\right)\quad\quad \Gamma_{\rm
\phi}=\frac{\Gamma_{\rm r}}{2} +2\pi\alpha k_{\rm
B}\frac{\epsilon^2}{E_{\rm eff}} T/\hbar.
\label{gphi}
\end{equation}
Note, that there may be nonexponential contributions to the dynamics
as well \cite{Loss}.

It is important to notice that in the strongly entangled case,
$\Delta_{\rm eff}\ll\Delta$, the relaxation rate, which describes the
thermalization of the system independent from the initial state, is
strongly reduced, whereas the dephasing rate, which describes the
projection of a superposition into a mixture of the eigenstates
$H_{\rm eff}$  is hardly affected. This is a very favorable situation
for a practical  measurement: The information is quickly available,
after $\tau_{\rm\phi} =\Gamma_{\rm\phi}^{-1}$ and is destroyed only
after $\tau_{\rm R}= \Gamma_{\rm R}^{-1}$. This is not only convenient
for experimental  implementation but also guarantees high fidelity:
The probability for reading out the correct result after the dephasing
time is   $P=e^{-\tau_{\rm \phi}/\tau_{\rm R}}$ and thus close to
unity.  For completing the description of the measurement, one has to
evaluate the resolution of the detector and the typical measurement
times. This depends on details of the physical realization 
of interest and has been done
in Ref.\ \cite{Neumann} for the superconducting setup.

\section{Summary and outlook}

We have studied the spin boson model with a structured bath using 
three different approaches: Bloch-Redfield, NIBA, and flow equation
renormalization.  We have arrived at a number of common features: If
the peak in the spectral  density is at frequencies much above the
environmental resonance, the  system is equivalent to a renormalized
Ohmic spin boson model. This has been interpreted in terms of quantum
measurement and the usefulness of this result for modeling quantum
detectors has been outlined. We have  furthermore shown that the
tunneling matrix element of the spin part  is renormalized downward if its
initial value $\Delta$ is below the environmental resonance $\Omega$ 
and renormaized upward
if it is above. We have compared this renormalization from all
approaches and shown that they are in  reasonable agreement within the
scope of their applicability. In particular, our analytical results
from NIBA and Bloch-Redfield reliably approximate the numerical
results from flow equations.

This work would not have been possible without the extensive,
illuminating  and thorough work on the Spin Boson model by Prof.\ Uli
Weiss. We would  like to express our high regards and respect for him
and his work and to congratulate him on his 60th birthday.

We would like to thank M.\ Grifoni and S.\ Kehrein for useful
discussions.  Work supported by the ARO under contract Nr.\
P-43385-PH-QC and through SFB 631.



\appendix
\section{Calculation of rates including poles of the spectral density}

We now want to outline how to calculate the rates
eq. \ref{eq:br_basicrate} We can interchange the order of integration
and evaluate the time integral, which can be expanded into a delta
function contribution and a Cauchy principal value. We can split
$\Gamma(E)$ into real and imaginary part, $\Gamma^\prime (E)$ and
$\Gamma^{\prime\prime}(E)$ and find
\begin{equation}
\Gamma^\prime(E)=\frac{1}{8\hbar}J(E)\left[\coth(\beta E/2)-1\right]
\end{equation}
for the real part, which determines the decoherence and
\begin{equation}
\Gamma^\prime(E)=\frac{1}{4\pi\hbar}\int d\omega\;  J(\omega){\mathcal
P}\frac{1}{\omega^2-E^2}\left[\coth(\beta \omega/2)E-\omega\right]
\label{eq:br_imagrate}
\end{equation}
for the imaginary part, which controls the frequency shifts. The
latter can be calculated by extending the integration contour to the
complete real axis, applying the residue theorem and resumming the
resulting Matsubara  series. We end up with
\begin{equation}
\Gamma=^{\prime\prime}\frac{\alpha}{2\pi}\frac{\Omega^2 E}{2i\Gamma} \sum
\frac{\sigma}{E^2-(\sigma\Omega+i\Gamma)^2}\left[p(\Gamma-i\sigma)-
{\rm Re}p(iE)-\pi\frac{\Gamma-i\sigma \Omega}{E}\right]
\label{eq:br_imgamma}
\end{equation}
where $p(x)=\psi(1+\beta x/2\pi)+\psi(\beta x/2\pi)$ involves the
digamma function $\psi$.

\end{document}